# The decline in the concentration of citations, 1900–2007

Vincent Larivière, Yves Gingras

Observatoire des sciences et des technologies (OST), Centre interuniversitaire de recherche sur la science et la technologie (CIRST), Université du Québec à Montréal, Case Postale 8888, succ. Centre-Ville, Montréal (Québec), H3C 3P8, Canada. E-mail: lariviere.vincent@uqam.ca; gingras.yves@uqam.ca

Éric Archambault

Science-Metrix, 1335A avenue du Mont-Royal E, Montréal, Québec, H2J 1Y6, Canada and Observatoire des sciences et des technologies (OST), Centre interuniversitaire de recherche sur la science et la technologie (CIRST), Université du Québec à Montréal, Montréal (Québec), Canada. E-mail: eric.archambault@science-metrix.com

## Abstract

This paper challenges recent research (Evans, 2008) reporting that the concentration of cited scientific literature increases with the online availability of articles and journals. Using Thomson Reuters' Web of Science, the present paper analyses changes in the concentration of citations received (two- and five-year citation windows) by papers published between 1900 and 2005. Three measures of concentration are used: the percentage of papers that received at least one citation (cited papers); the percentage of papers needed to account for 20%, 50% and 80% of the citations; and, the Herfindahl-Hirschman index. These measures are used for four broad disciplines: natural sciences and engineering, medical fields, social sciences, and the humanities. All these measures converge and show that, contrary to what was reported by Evans, the dispersion of citations is actually increasing.

## Introduction

In a recent paper, Evans (2008) challenged commonly held beliefs about online availability of journals and papers by showing that an increase in their online availability and their historical archives 1) decreased the age of cited scientific literature and 2) increased the concentration of citations on a smaller proportion of published papers. In other words, though more research (older and recent) is now available online, researchers cite more recent papers and concentrate their citations on fewer papers. As Evans puts it, the online availability of scientific papers and journals leads researchers to "…weave into a more focused—and more narrow—past and present." (p. 398)

Evans' claims on the younger age of cited literature are contradicted by empirical studies that show that researchers cite an increasingly older body of scientific literature (Larivière, Archambault and Gingras, 2008), an observation that is backed by both theory (Egghe, 1993, 2008; Glänzel and Schoepflin, 1994, 1995) and studies on researchers' patterns of use (e.g. Tenopir and King, 2008). Evans' assertion on the increasing concentration of citations reflects a widely held belief (Hamilton 1990; 1991) that most scientific articles are never cited, a common lore that comes back periodically in the literature (e.g. Meho, 2008; Macdonald and Kam, 2007). Though several empirical studies have challenged this belief (Abt, 1991; Garfield, 1998; Pendlebury, 1991; Schwartz, 1997, Stern, 1990, Van Dalen and Henkens, 2004), no study has as yet measured the changes in the proportion of cited/uncited articles over a long period of time. As suggested by Pendlebury (1991), "[a] trend toward more or less "uncitedness," however, might be meaningful. For the 1980s, we see no such trend in the scientific literature: the numbers are essentialy flat …" (p. 1410.)

Through a detailed analysis of citations to publications over the 1900–2007 period, the present paper shows very clearly that the proportion of uncited papers and the concentration of citations received are *decreasing* rather than increasing. The next section of this paper briefly presents the methods and database used and is followed by a presentation of the results obtained. The last section compares our results with those of Evans (2008).

**Methods**

Three measures of the concentration of citations received by scientific papers are presented. The first is the percentage of papers published in a given year that received at least one citation two and five years after publication (cited papers). This means that complete citation windows end in 2005 for the two-year window and in 2002 for the five-year window (including publication year). The higher the proportion of cited paper is, the more citations are dispersed across a large percentage of published papers and, hence, the smaller the concentration.

The second indicator of citation concentration is the percentage of papers needed to account for 20%, 50% and 80% of the total citations received by papers published in a given year. If, over the years, a smaller percentage of the top papers are needed to account for each percentage of the citations, then the concentration is increasing. If a higher percentage of papers is needed to account for each percentage, then the concentration is decreasing. Unlike analyses of *references made*, where uncited papers are *de facto* excluded, or other analyses of the distribution of citations received (Price, 1976; Lehman, Lautrup and Jackson, 2003), uncited papers are included in our analysis of the concentration of the distribution of citations. This is an important advantage of using *citations received* instead of *references made* (Price, 1963).

The third and final measure of concentration presented in this paper is the Herfindahl-Hirschman index (HHI), a measure of the concentration of firms in a given market often used by antitrust authorities in the U.S. It can be simply defined as the sum of squares of firms' market share: the higher the HHI, the more concentrated the market is. This is the sole indicator used by Evans (2008) to measure the concentration of citations. When applied to citations, we consider the size of the market to be the sum of the number of citations received by each individual paper, and the market shares to be the number of citations received by each paper divided by the total number of citations received by papers published the same year. Hence, if papers published in 2000 received a total of 20 million citations, the market share of each paper is its number of citations received divided by 20 million. The market share of each paper is then squared and the results are summed to obtain the

HHI of papers published in 2000. Given that, by definition, uncited papers do not have any *market share*, they are *de facto* excluded from the calculation of this index.

Data for this paper are drawn from Thomson Scientific's Web of Science, which comprises the Science Citation Index Expanded (SCIE), Social Sciences Citation Index (SSCI) and Arts and Humanities Citation Index (AHCI), for the 1900–2007 period. Each journal was classified based on the taxonomy used by the U.S. National Science Foundation. For the Humanities, the NSF classification was completed using in-house classification results. NSF subject headings where grouped into four broad categories: natural sciences and engineering (NSE), medical fields (MED), social sciences (SS), and the humanities (HUM). Data for NSE and MED start in 1900, data for the SS start in 1956 and for HUM in 1975.

The matching of article citations was made using Thomson's reference identifier provided with the data, as well as using the author, publication year, volume number and page numbers. Only citations received by articles, notes and review articles were included in the study and first author self-citations were excluded. On the whole, citations received by a total of more than 27 million papers (11 million papers in NSE, 12.7 million in MED, 2.5 million in SS and 0.9 million in HUM) are retrieved in a pool of more than 615 million references contained in the database.

**Results**

Figure 1 shows that the percentage of papers that received at least one citation two and five years after publication increased steadily throughout the period, except between 1960 and 1970. Indeed, whereas citations received were concentrated on 10% to 20 % of published papers at the beginning of the last century and on about half of all papers at the beginning of the Seventies, in 2005, the last year for which we have a complete two-year citation window, citations were distributed among 80% of published papers in MED, 60% of papers in NSE and 55% of papers in SS. When one uses a five-year citation window, the general trends are the same, and only 12% of papers in MED, 27% in NSE and 32% in SS remained uncited in 2002. Though not shown, data using a ten-year citation window follow the same trend, albeit with even higher rates of citedness.

In fact, only the broad field of HUM behaves differently, as it does with regard to several other aspects of scholarly communication, such as collaboration (Larivière, Gingras and Archambault, 2006) and the use of serials (Larivière, Archambault, Gingras and Vignola-Gagné, 2006). The very low percentage of articles cited at least once may be a reflection of the tendency of humanities researchers to cite books instead of articles. All in all, these data strongly show that, in all fields except HUM, fewer and fewer of the published papers go unnoticed and uncited and, consequently, science is increasingly drawing on the stock of published papers.

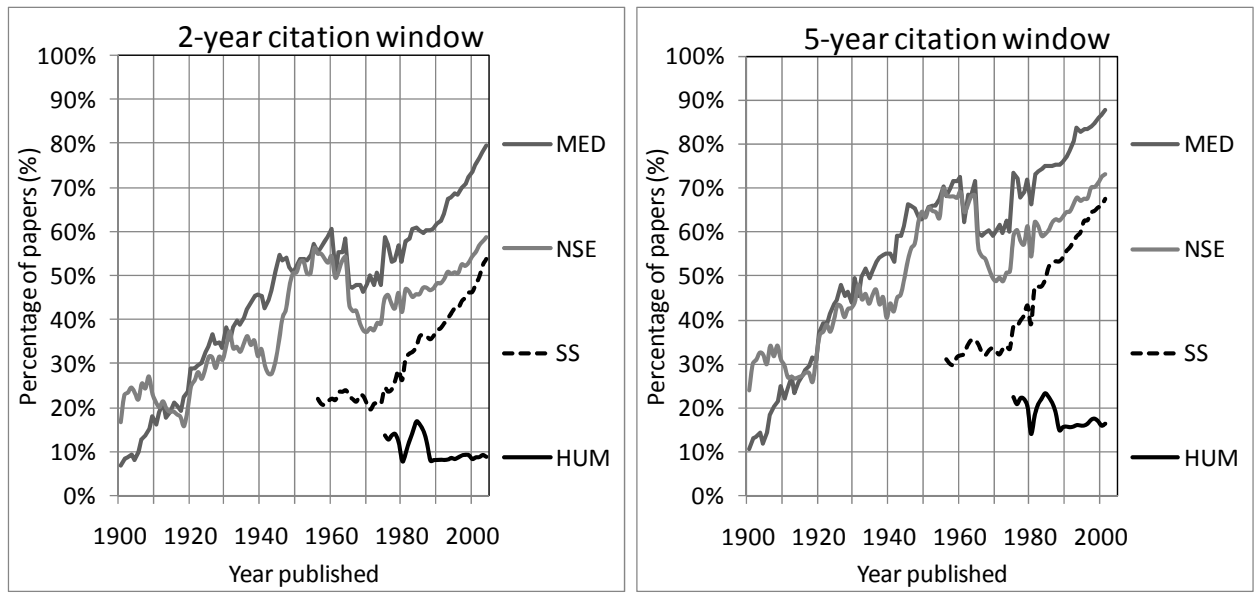

Figure 1. Percentage of papers that received at least one citation, two- and five-year citation windows, by field, 1900–2005 and 1900–2002

Figure 2 presents the percentage of published papers needed to account for the top 20%, 50% and 80% of citations received two years after publication. NSE and MED follow a similar pattern: citations were increasingly dispersed from the beginning of the last century until the Sixties, when they started to become increasingly concentrated among a smaller proportion of published papers. This phase of increased concentration ended around 1990 and, since then, the dispersion of citations received has steadily increased. For instance, in 2005, 33% of MED papers and 28% of NSE papers accounted for 80% of the citations received, compared to respectively 24% and 23% in 1990. In SS, the dispersion of citations has been increasing continuously since 1956, and at an even faster rate since 1990. In 2005, 28% of the papers accounted for 80% of the citations, compared with 19% in 1990 and 14% in 1956. These empirical data suggest that there may be an approximate 15-year lag following a growth or decrease in the number of papers published during which the concentration of citations falls (for the historical growth rate of publications in these fields, see Larivière, Archambault and Gingras, 2008).

As one would expect from HUM data in Figure 1—which shows that citations received were concentrated on a very small share of the papers and that the trend was flat—an extremely small percentage of papers account for the majority of citations. Indeed, in 2005, 0.5% of papers accounted for 20% of citations, 2.6% for 50% of citations and 7.2% of papers for 80% of citations received. Apart from a small "bump" in the data, which can very likely be attributed to the poor quality of the data in HUM at the beginning of the Eighties, no trend can be discerned. The extremely skewed nature of the data in HUM, again, suggests that extreme caution should be applied in using journal-based bibliometric data for the evaluation of research in HUM.

Hence, for NSE, MED and SS, the dispersion of citation has been mostly *increasing* since the beginning of the 20th century. Although the distributions of citations received are still highly concentrated and a minority of papers still account for a majority of the citations, this level of concentration has been decreasing over time. Moreover, in MED and in SS, citations received by papers published in 2005 had the lowest concentration in history. These data thus clearly show that, contrary to Evans' findings (2008), the concentration has been decreasing over time in these three broad fields and that citations received are *increasingly dispersed* among a larger percentage of published papers, instead of being more concentrated as time goes on, as suggested by Evans.

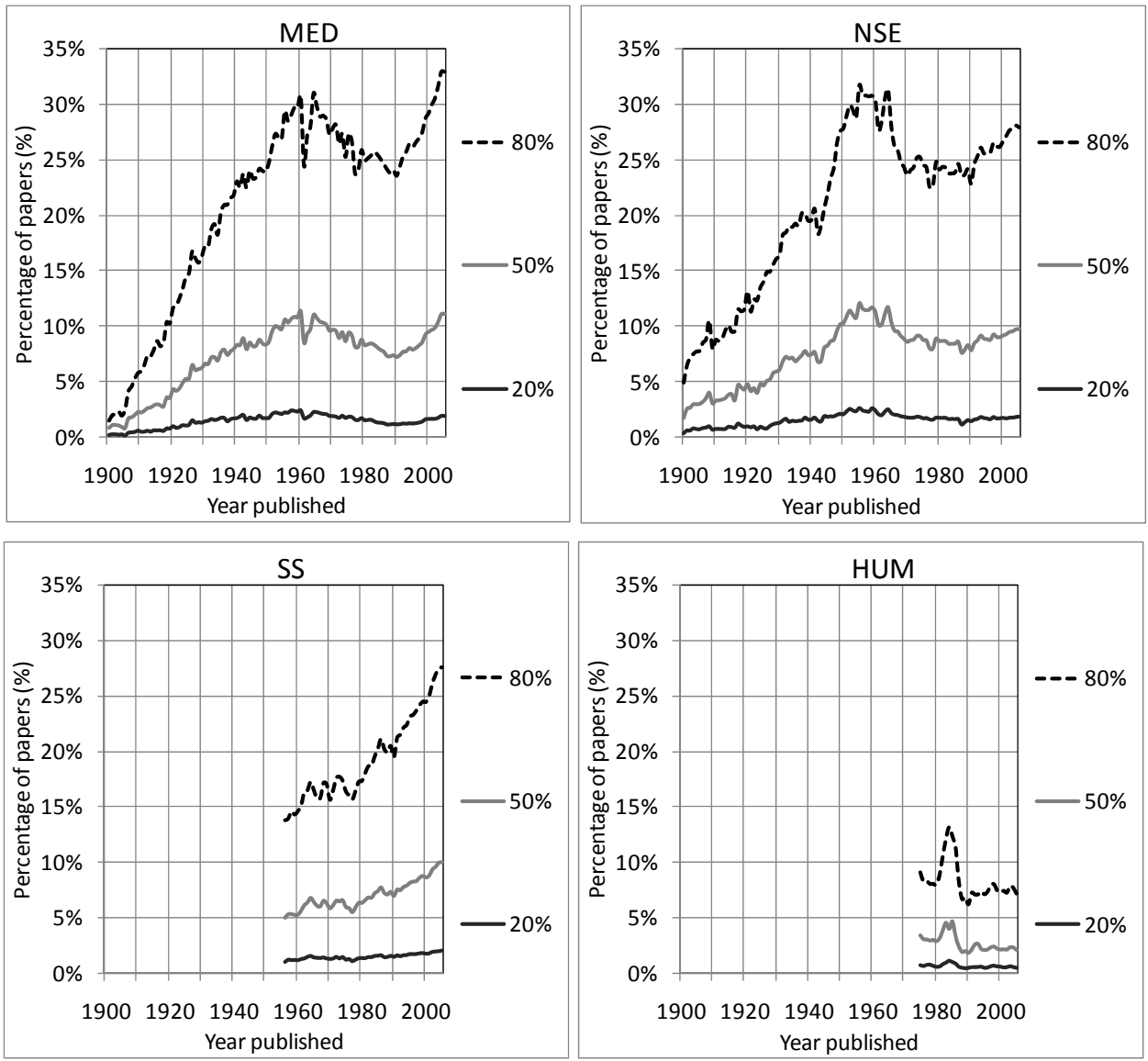

Figure 2. Percentage of papers needed to obtain 20%, 50% and 80% of the citations received using a two-year citation window, by field, 1900–2005

One could argue, however, that we have not used the same measure of concentration as that used by Evans (2008). Figure 3 accordingly shows the evolution of the HHI for citations received two and five years after publication. One can readily see that, as could be expected given the foregoing results, the concentration of citations received has also decreased considerably since the beginning of last century, a result that simply reflects the exponential increase in the number of papers published and cited. One can also see that, in MED and NSE, citations received became more concentrated during the two World Wars. Given that fewer papers were published *during* the wars, researchers chose their references among a smaller pool of papers.[1] This had the effect of diminishing the HHI, which is highly sensitive to the number of "competing" units. But what is even more important is that, in contrast to what Evans (2008) reported using the same index, the HHI of citations received steadily decreased over the period studied, except during the two World Wars and, for a brief period, at the end of the Eighties. Hence, for all fields except HUM, papers published in 2005 had the lowest concentration of citations received in history. Though it is not shown, we have also compiled the HHI values from the point of view of references made to papers as well as to journals. The tendency is *exactly the same*; and 2007 is the year in which references made were the least concentrated.

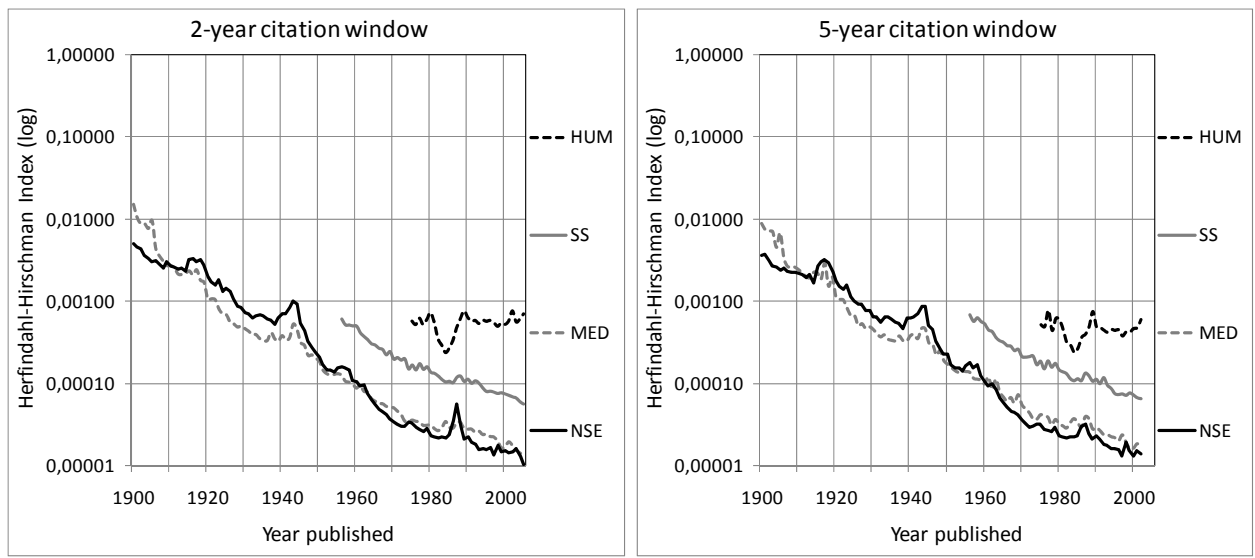

Figure 3. Herfindahl-Hirschman index of citations received, two and five year citation window, by field, 1900–2005 and 1900–2002

**Discussion and Conclusion**

Because of the multiple measures used and the clearly documented method associated with the simplicity of the protocol used here, the present paper provides clear and practically irrefutable evidence that, at the macro level, the concentration of *citations received* has been decreasing in NSE, MED and SS. First, the percentage of papers which received at least one citation has been increasing since the Seventies. Second, the percentage of papers needed to account for 20%, 50% and 80% of the citations received has been increasing, and third, the HHI has been steadily decreasing since the beginning of the last century. All these measure converge to demonstrate that citations are not becoming more concentrated but increasingly dispersed, and one can therefore argue that the scientific system is increasingly efficient at using published knowledge. Moreover, what our data shows is not a tendency towards an increasingly exclusive and elitist scientific system, but rather one that is increasingly democratic.

---

[1] As shown by Larivière, Archambault and Gingras (2008), this had the effect of *increasing* the age of cited literature.

The data reported in this paper do not take into account the "online availability" variable. Hence, it does not provide direct proof that the online availability of articles is not negatively correlated with an increased concentration of citations received by articles, nor can it prove that electronic publishing and access drives the tendencies observed. However, given that 1) most journals are available online and 2) the phenomenon observed by Evans (2008) is not observed at all at the macro level—in fact the opposite can be observed—it is either a marginal phenomenon or an artefact. A possible explanation of these results is that, in measuring the age of cited literature, Evans failed to use any clearly defined interval between the "breadth" of what was available in a given year and the age of materials cited; this would undoubtedly have an effect on the age of what is being cited. In order to derive a relation that takes into account the delays between finding, reading, citing and publishing a paper, one should correlate the age of what is cited with what was published a given number of years before.

In conclusion, our own extensive investigations on this phenomenon, presented here and previously (Larivière, Archambault and Gingras, 2008), show that Evans' suggestions that researchers tend to concentrate on more recent and more cited papers does not hold at the aggregate level in the biomedical sciences, the natural sciences and engineering, or the social sciences. Though many factors certainly contribute to the observed trends, two things are clear: researchers are not increasingly relying on recent science, nor are citations limited to fewer papers or journals.

**Acknowledgements**

The authors wish to thank Alain Couillard, Stevan Harnad, Johanna Kratz, Jean-Pierre Robitaille and Jillian Tomm for their valuable comments and suggestions.